\documentclass[journal]{IEEEtran}

\usepackage{graphicx}
\graphicspath{ {./} }

\usepackage{array}
\usepackage{caption}
\usepackage{float}
\usepackage[caption = false]{subfig}

\usepackage[dvipsnames]{xcolor}

\usepackage{multirow}
\usepackage{amsmath}

\ifCLASSINFOpdf
\else
\fi

\begin{document}
%

\title{Inferior Myocardial Infarction Detection from lead II of ECG: A Gramian
Angular Field-based 2D-CNN Approach }

%
%
%

\author{
\IEEEauthorblockN{
Asim Yousuf\IEEEauthorrefmark{1}, Rehan Hafiz\IEEEauthorrefmark{1}, Saqib Riaz\IEEEauthorrefmark{2}, Muhammad Farooq\IEEEauthorrefmark{3}, Kashif Riaz\IEEEauthorrefmark{1}, M. Mahboob Ur Rahman\IEEEauthorrefmark{1} }

\IEEEauthorblockA{\IEEEauthorrefmark{1} Electrical engineering department, Information Technology University, Lahore 54000, Pakistan\\ \IEEEauthorrefmark{2} Bahawal Victoria Hospital, Bahawalpur, Pakistan \\
\IEEEauthorrefmark{3} Phyn LLC, Torrance, CA, USA \\ 
\IEEEauthorrefmark{1}\{kashif.riaz, mahboob.rahman\}@itu.edu.pk, \IEEEauthorrefmark{3}mfarooq@crimson.ua.edu \\ }
}

\maketitle


\begin{abstract}

This paper presents a novel method for inferior myocardial infarction (MI) detection using lead II of electrocar-
diogram (ECG). We evaluate our proposed method on a public dataset, namely, Physikalisch Technische Bundesanstalt
(PTB) ECG dataset from Physionet. Under our proposed method, we first clean the noisy ECG signals using db4 wavelet,
followed by an R-peak detection algorithm to segment the ECG signals into beats. We then translate the ECG timeseries
dataset to an equivalent dataset of gray-scale images using Gramian Angular Summation Field (GASF) and Gramian
Angular Difference Field (GADF) operations. Subsequently, the gray-scale images are fed into a custom two-dimensional
convolutional neural network (2D-CNN) which efficiently differentiates between a healthy subject and a subject with MI. Our
proposed approach achieves an average classification accuracy of 99.68
dataset with noise and baseline wander, GADF dataset with noise and baseline wander, GASF dataset with noise and
baseline wander removed, and GADF dataset with noise and baseline wander removed, respectively. Most importantly,
this work opens the floor for innovation in wearable devices to measure lead II ECG (e.g., by a smart watch worn on right
wrist, along with a smart patch on left leg), in order to do accurate, real-time and early detection of inferior wall MI.

\end{abstract}

\begin{IEEEkeywords}

Cardiovascular Diseases, Myocardial Infarction, Electrocardiogram, Convolutional Neural Network, Gramian Angular Difference Field, Gramian Angular Summation Field.

\end{IEEEkeywords}

%

\section{Introduction}
%
%
%
%

According to the world health organization (WHO), cardiovascular diseases (CVDs) are among the leading cause of death in the people worldwide \cite{who1}. Myocardial infarction (MI), more commonly known as heart attack, is a frequently occurring CVD. A very brief pathophysiology of the MI is as follows. MI is often caused by the coronary artery disease (coronary arteries supply the Oxygen-rich blood to the heart muscle known as myocardium, after each contraction). Occasionally, the blood supply to myocardium is reduced due to the build-up of cholesterol-rich plaques inside the walls of one of the coronary arteries, leading to what is called cardiac ischemia. If it remains untreated, cardiac ischemia may lead to MI \cite{acharya2}, \cite{heart3}. MI is a life-threatening condition, and if clinical intervention is not made within hours (or days) of onset, it may result in injury and irreversible damage of the heart muscle leading to a heart attack \cite{k-a4}. MI is also known as a silent heart attack because many times the heart muscle starts to die while patients are not even aware of it. It has been estimated that among all the heart attacks that are reported, around 72\% are the silent attacks \cite{mozaffarian5}.   

Electrocardiogram (ECG), a cost-effective and non-invasive method, is unanimously considered as a golden standard for diagnosing a wide range of cardiac irregularities \cite{acharya2}. Thus, ECG could provide viable hints/cues about the MI at early stages \cite{acharya6}. An ECG signal comprises 12 leads including lead I, II, III (standard leads), V1-V6 (chest leads), and aVL, aVR, aVF (limb leads). For the detection of MI, cardiologists need to analyze manually the ECG traces of the affected individuals, which is a time-consuming and error-prone method. Therefore, an artificial intelligence (AI)-aided automatic MI detection system is the need of the hour.

The conventional machine learning (ML) techniques have shown great promise for semi-automatic MI detection with good accuracy. Some prominent techniques used by the ML methods include the following: wavelet decomposition \cite{raghavendra7}, sparse decomposition \cite{raj10}, linear discriminant analysis \cite{chazal11}, and support vector machines (SVMs) \cite{osowski12}, multi-scale principal component analysis \cite{alickovic13}, wavelet packet decomposition \cite{li15}, and independent component analysis \cite{elhaj14}. 

All of the above-mentioned ML methods have some drawbacks. First, such methods rely upon extraction of handcrafted features by the domain experts. Besides, it is difficult to retain the generalization capabilities of the ML classifiers on the unseen data because the classification results may rely on the nature of handcrafted features which could change dramatically with the change in age, gender, or medical history of the new patients. Deep learning (DL) techniques, on the other hand, have the promise of automated feature extraction \cite{ince17}, \cite{jun18}, \cite{acharya19}, \cite{tan20}. Thus, inline with the most recent DL-based works, this work proposes a novel DL-based method for automatic MI detection.

{\bf Contributions:} 
The main contribution of this work is two-fold. 1) The proposed method utilizes the Gramian angular field (GAF) method in order to convert the one-dimensional ECG timeseries (with and without noise) from a public dataset into two-dimensional gray-scale images. 2) The GAF images are fed into a custom 2D-convolutional neural network (CNN) model that automatically extracts the complex representative features from the 2D input data and classifies these images as either belonging to healthy or MI subjects.  

{\bf Outline:} 
The rest of this paper is organized as follows. Section II summarizes the selected related work. Section III describes the key statistics of the public ECG dataset used in this work, key data pre-processing steps, Gramian Angular field-based transformation of 1D ECG signal to a 2D image, and the architecture of the custom 2D-CNN. Section IV outlines performance metrics, provides pertinent details regarding experimental setup, and discusses selected results. Section V concludes the paper.

\section{Related Work} 

Several researchers have employed DL techniques for the identification or localization of ECG arrhythmias.
Authors in \cite{kiranyaz21} used 48 records of lead-II ECG, and proposed a 1D-CNN model for classification of normal and four different arrhythmias using ECG signals. 
\cite{mostayed22} used 6,877 records of 12-leads ECG, and implemented a bi-directional long-short-term-memory (LSTM) network to classify the ECG signals into normal and eight different arrhythmias. 
Authors in \cite{zhang23} the 48 records of lead-II ECG, and presented an R-peak detection method and a recurrent neural network (RNN). Their proposed method learns time correlation from lead-II ECG of 48 records and classifies ECG signals into normal and four different arrhythmias. 
Authors in \cite{li24} used 549 records of lead-II ECG, and implemented a 1D-CNN for the classification of normal ECG and eight different arrhythmias.  
Authors in \cite{acharya25} utilized 106,501 ECG beats as $128\times 128$ gray-scale images, and proposed an R-peak detection method and a 1D-CNN for the classification of normal and MI beats.  
Authors in \cite{jun26} proposed a method based on pattern recognition and a 2D-CNN to classify between a normal ECG and seven different arrhythmias. 
\cite{xu27} used 108,655 beats of lead-II ECG, and implemented a deep neural network to identify normal and four different arrhythmias. 
Finally, authors in \cite{huang28} used 2,520 ECG records as a $256 \times 256$ images, and proposed a novel approach using short time-frequency transform-based spectrograms and a 2D-CNN for the classification of normal ECG and seven different arrhythmias.

\section{Methodology}

This section describes the key details of the proposed automated MI detection system that utilizes the Gramian Angular Field operation and a 2D-CNN. The main components of the proposed method are illustrated in Fig. 1, and the pertinent details for each of the component are discussed below.  

\subsection{Physikalisch-Technische Bundesanstalt Dataset}

We utilize Physikalisch-Technische Bundesanstalt (PTB) ECG dataset \cite{physionet29} from Physionet. The dataset consists of noisy ECG data of 148 MI subjects: 110 men and 38 women (aged 36 to 86). The dataset also consists of noisy ECG data of 52 healthy subjects: 39 men and 13 women (aged 17 to 81). Each ECG trace is 100 seconds long, recorded at 1000 samples per second. Table I summarizes the key details of the PTB dataset. For each subject, 12-lead ECG data is provided, but we utilize only the lead-II of the ECG data in this work. Furthermore, to validate the hypothesis that our proposed classifier is robust to noise, we prepare two types of datasets. The first dataset consists of the raw and noisy (lead-II) ECG signals from PTB dataset as is, while we construct the second dataset by removing the noise and baseline wander from the ECG signals in the first dataset.

\begin{figure*}
  \includegraphics[width=\textwidth,height=7cm]{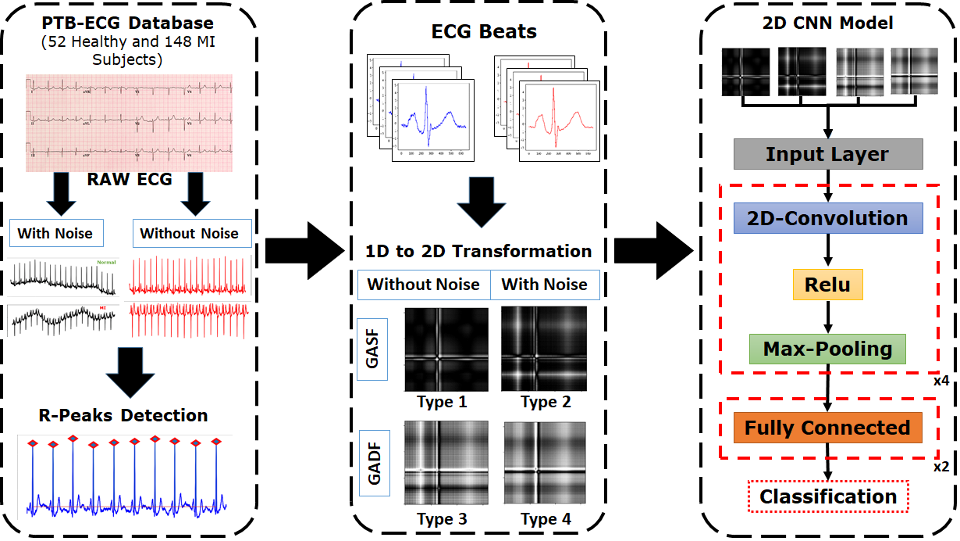}
  \caption{The proposed method for MI detection.}
\end{figure*}

\begin{figure}[h]
    \centering
    \includegraphics[width=0.5\textwidth]{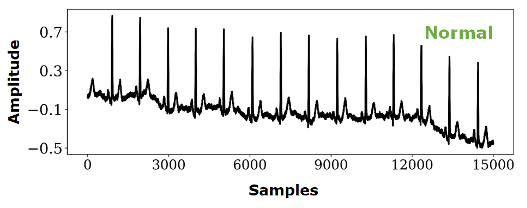}
    \caption{Lead-II ECG signal with noise and baseline wander (healthy class).  }
    \label{fig:mesh1}
\end{figure}

\begin{figure}[h]
    \centering
    \includegraphics[width=0.5\textwidth]{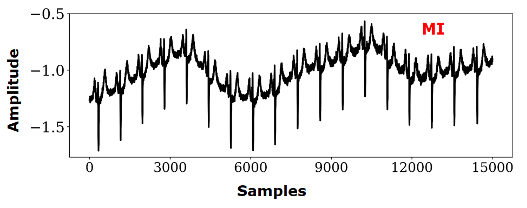}
    \caption{Lead-II ECG signal with noise and baseline wander (MI class). }
    \label{fig:mesh1}
\end{figure}

\subsection{Data Pre-processing}

Lead-II ECG signals for healthy and MI subjects with noise and baseline wander are shown in Fig. 2 and Fig. 3 respectively. In this work, Daubechies 4 wavelet is utilized to remove high-frequency noise and baseline wander from the ECG signals \cite{singh30}. Thus, clean ECG signals for healthy and MI subjects (after denoising) are shown in Fig. 4 and Fig. 5, respectively. The R peaks (positive peak in the QRS complex) are detected in both datasets (noisy and clean) by means of Pan-Tompkins algorithm \cite{pan31}. An ECG trace with its R-peaks detected is shown in Fig. 6.

\begin{figure}[h]
    \centering
    \includegraphics[width=0.5\textwidth]{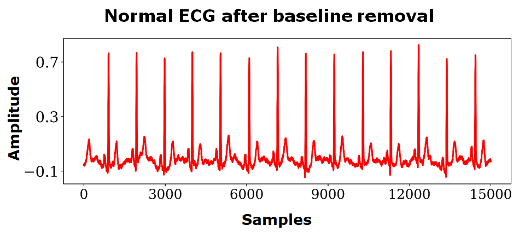}
    \caption{Lead-II ECG signal with noise and baseline wander removed (healthy class). }
    \label{fig:mesh1}
\end{figure}

\begin{figure}[h]
    \centering
    \includegraphics[width=0.5\textwidth]{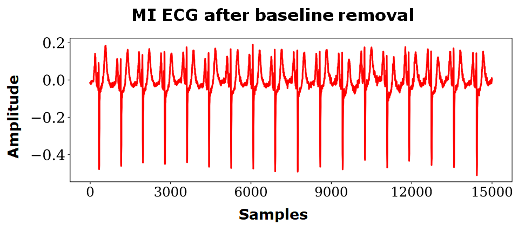}
    \caption{Lead-II ECG signal with noise and baseline wander removed (MI class).  }
    \label{fig:mesh1}
\end{figure}

\begin{figure}[h]
    \centering
    \includegraphics[width=0.5\textwidth]{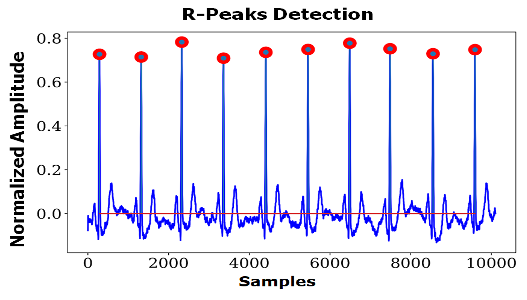}
    \caption{An ECG trace with its R-peaks detected.  }
    \label{fig:mesh1}
\end{figure}

The Pan-Tompkins algorithm allows us to segment the ECG signal into beats, for both datasets. All ECG signals are segmented in such a way that 250 samples are considered before the R-peak index to involve P-peak, while 400 samples are considered after R-peak to involve T-peak, in order to form one beat. Thus, each beat comprises a window of 651 samples that together capture one P-QRS-T waveform. The segmentation allowed us to extract 40,267 beats in total, for both datasets (noisy and clean). We then applied Z-score normalization to remove the offset effect and address the issue of amplitude scaling \cite{jain32}. Fig. 7 and Fig. 8 show the extracted time-domain P-QRS-T wave (basically, 1 cardiac beat) for the healthy and MI subjects, respectively. 

\begin{figure}[h]
    \centering
    \includegraphics[width=0.5\textwidth]{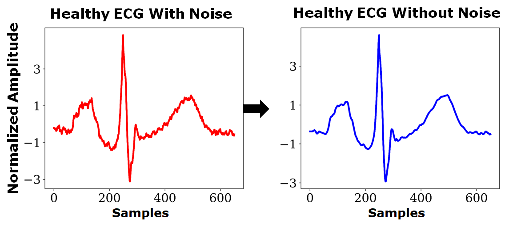}
    \caption{One ECG beat of a healthy subject.  }
    \label{fig:mesh1}
\end{figure}

\begin{figure}[h]
    \centering
    \includegraphics[width=0.5\textwidth]{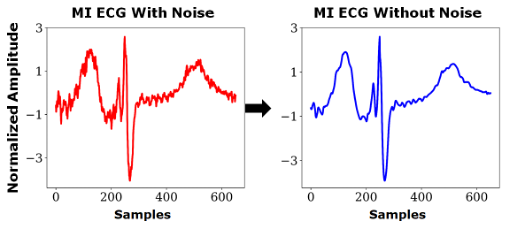}
    \caption{One ECG beat of an MI subject. }
    \label{fig:mesh1}
\end{figure}

\begin{table}[]
\centering
\begin{tabular}{|l|l|l|l|l|p{8mm}|}  \hline
Type    & Males & Females & Min Age & Max Age & No: of Beats \\ \hline 
Healthy & 39          & 13            & 17          & 81          & 10,139       \\ \hline
MI      & 110         & 38            & 36          & 86          & 30,128       \\ \hline
Total   & 149         & 51            & -           & -           & 40,267      \\ \hline
\end{tabular}
\caption{Key details of the PTB dataset}
\end{table}

\subsection{Gramian Angular Field Method}

Our custom 2D-CNN requires images (i.e., two-dimensional data) on its input in order to differentiate between the healthy and MI subjects. Therefore, each ECG beat is encoded as an image using the Gramian angular field (GAF) operator. Two kinds of GAFs, i.e., Gramian angular field with summation (GASF) and Gramian angular field with difference (GADF) are used to transform each time-domain ECG beat to an image as follows. First, a normalized time-domain ECG beat consisting of $n$ samples in Cartesian coordinates is converted into a polar coordinate system. Next, a Gramian matrix of dimension $n\times n$ is formed using trigonometric cosine functions. Specifically, GASF method computes the cosine of sum of two angles to compute each element of the Gramian matrix, while GADF computes the sine of subraction of two angles to compute each element of the Gramian matrix. 

More precisely, the 1D ECG timeseries is converted to the 2D image as follows.
Let $X=\{x_1,x_2,…,x_n\}$ represent the ECG beat timeseries consisting of $n$ real-valued samples. Then, the first step is to re-scale the $i$-th sample $x_i$ of the ECG beat as follows:
\begin{align}
    xnor_i = \frac{(x_i-min(X))+(x_i-max(X))}{max(X)-min(X)}
\end{align}
Thus, the Min-Max scaler normalizes each ECG beat such that its amplitudes now lie in the range $[-1,1]$. The next step converts the ECG beat (from Cartesian) to polar coordinates as follows:
\begin{align}
    \emptyset_i = \cos^{-1}(xnor_i) \; \; \; -1\leq xnor_i \leq 1
\end{align}
It is easy to verify that the polar coordinate $\emptyset_i$ of the $i$-th sample $xnor_i$ of the normalized time-domain ECG beat falls in the angular range of $[0,\pi]$. An angular resolution in the range $[0,\pi]$ provides significant information to any classifier that utilizes images constructed by the GASF and GADF operations. 

The next step is to compute pair-wise trigonometric sum and difference (for every two polar-coordinate samples in the ECG beat) in order to capture the temporal correlation between the samples of the ECG beat at different time intervals. Accordingly, the Gram matrix corresponding to the GASF operation is given as follows: 

\begin{align}
    G_{GASF} = 
 \begin{pmatrix}
  \cos(\emptyset_1+\emptyset_1) & \cos(\emptyset_1+\emptyset_2) & \cdots & \cos(\emptyset_1+\emptyset_n) \\
  \cos(\emptyset_2+\emptyset_1) & \cos(\emptyset_2+\emptyset_2) & \cdots & \cos(\emptyset_2+\emptyset_n) \\
  \vdots  & \vdots  & \ddots & \vdots  \\
  \cos(\emptyset_n+\emptyset_1) & \cos(\emptyset_n+\emptyset_2) & \cdots & \cos(\emptyset_n+\emptyset_n) 
 \end{pmatrix}
\end{align}

Similarly, the Gram matrix corresponding to the GADF (trigonometric difference of sine) operation is as follows: 

\begin{align}
    G_{GADF} = 
 \begin{pmatrix}
  \sin(\emptyset_1-\emptyset_1) & \sin(\emptyset_1-\emptyset_2) & \cdots & \sin(\emptyset_1-\emptyset_n) \\
  \sin(\emptyset_2-\emptyset_1) & \sin(\emptyset_2-\emptyset_2) & \cdots & \sin(\emptyset_2-\emptyset_n) \\
  \vdots  & \vdots  & \ddots & \vdots  \\
  \sin(\emptyset_n-\emptyset_1) & \sin(\emptyset_n-\emptyset_2) & \cdots & \sin(\emptyset_n-\emptyset_n) 
 \end{pmatrix}
\end{align}
Note that $G_{GADF}$ is a hollow matrix with zeros on its main diagonal.

Finally, Fig. 9 provides a pictorial representation of the GAF method whereby the ECG beats time series (with and without noise) is transformed to images under the GASF and GADF operations. 

\begin{figure*}
  \includegraphics[width=\textwidth,height=6cm]{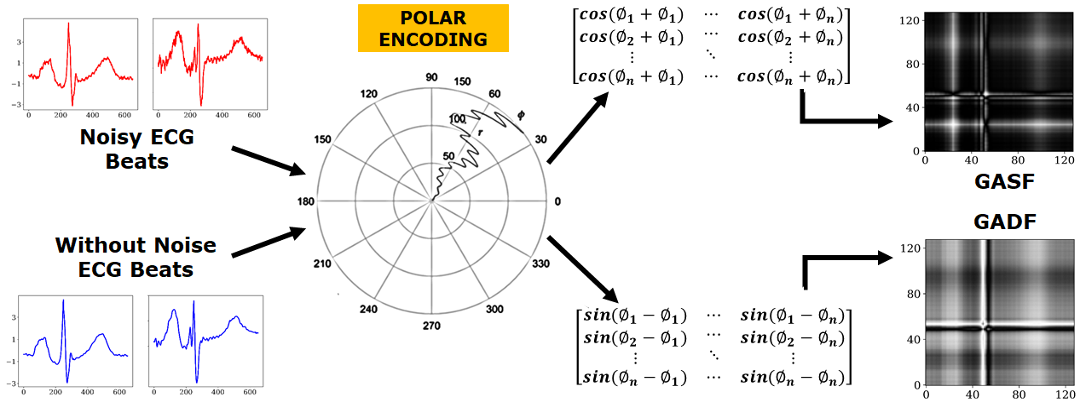}
  \caption{Illustration of ECG Beats timeseries encoded as images using Gramian Angular Field with summation and difference.}
\end{figure*}

To sum things up, each ECG beat consisting of 651 samples is transformed into an equivalent $128\times128$-pixel gray-scale image using the GAF method. This transformation preserves the temporal correlation between the samples of the 1D ECG beat signal. Fig. 10 and Fig. 11 show an example transformation of a sample ECG beat (with and without noise) into GASF and GADF images for a healthy and MI subject, respectively. We note that there is subtle difference between the GADF and GADF images for the normal and MI subjects. This attests to the feasibility of using GAF method for efficient classification (discrimination) between the two classes (healthy and MI).

\begin{figure*}
  \includegraphics[width=\textwidth,height=7cm]{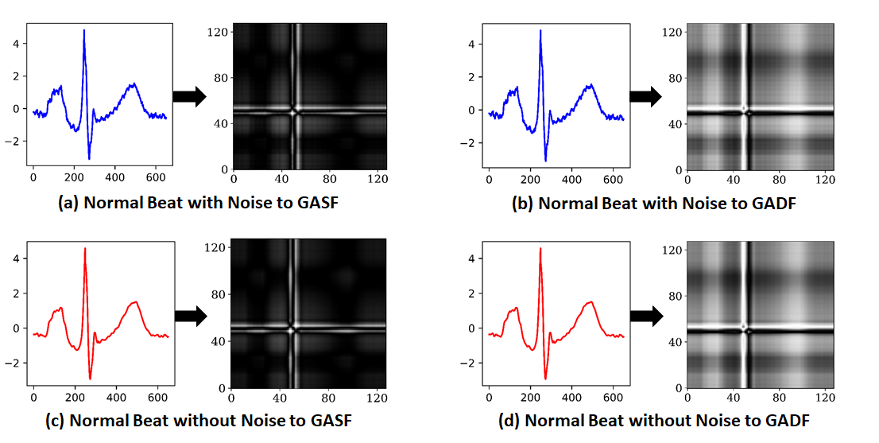}
  \caption{Healthy ECG Beats encoded as GASF and GADF images.}
\end{figure*}

\begin{figure*}
  \includegraphics[width=\textwidth,height=7cm]{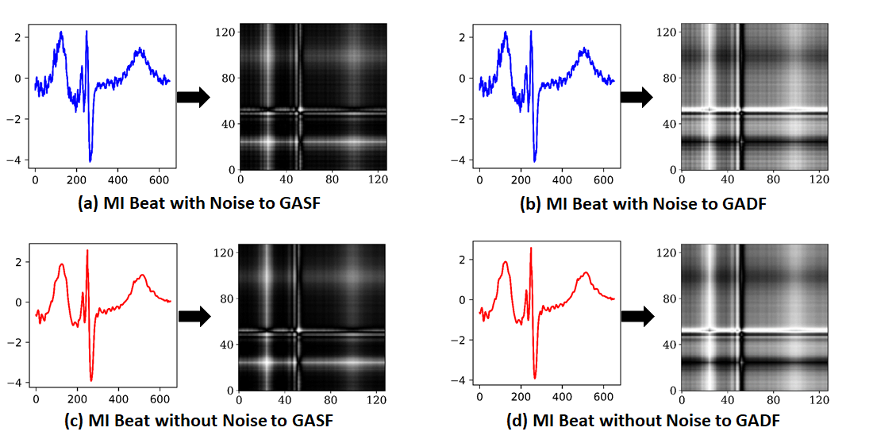}
  \caption{MI ECG Beats encoded as GASF and GADF images.}
\end{figure*}

\subsection{The Architecture of the Custom 2D-CNN}

We implement a custom multi-layer 2D-CNN for MI detection from GASF and GADF images. It is well known that the convolutional layers and pooling layers are important building blocks of a CNN. The convolutional layers do automatic feature extraction, while the pooling layers down-sample the input volume along its spatial dimension with care to retain the vital information. The customized 2D-CNN architecture in this work consists of four convolutional layers (with RELU activation function), followed by a max-pooling layer. Subsequently, the data is flattened and passed to the dense network with two fully-connected layers. A sigmoid activation function is used to do binary classification (of healthy and MI subjects). Fig. 12 shows the the block diagram of the customized 2D-CNN model, while Table II summarizes the architectural details of the custom 2D-CNN model used in this work.

\begin{figure*}
\centering
  \includegraphics[width=0.8\textwidth,height=6cm]{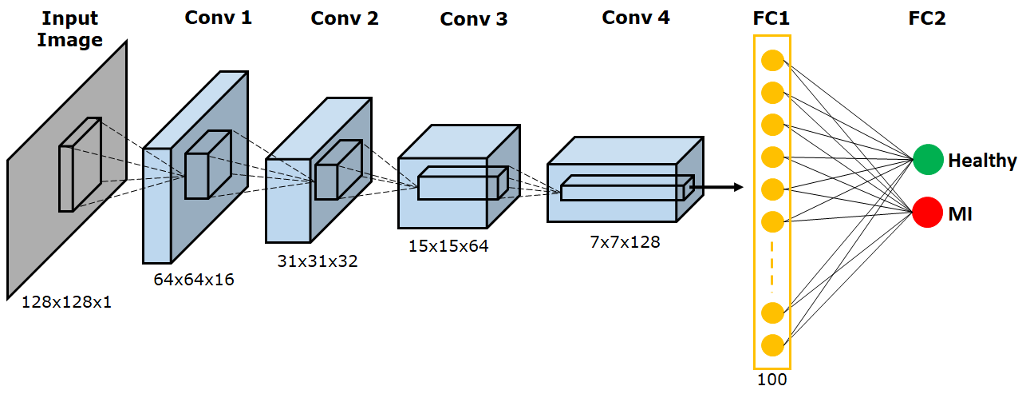}
  \caption{Block diagram of the custom 2D-CNN model used in this work.}
\end{figure*}

\begin{table*}[]
\centering

\begin{tabular}{|c| c| c| c| c| c|} \hline
    
Layers             & Layer Type                & No. of Kernels/Neurons & Stride & Input Shape    & Output Shape   \\\hline
\multirow{2}{*}{1} & Convolution + RELU        & 16                     & 1      & 128 x 128 x 1  & 128 x 128 x 16 \\ \cline{2-6}
                   & Max pooling               & -                      & 2      & 128 x 128 x 16 & 64 x 64 x 16   \\\hline
\multirow{2}{*}{2} & Convolution + RELU        & 32                     & 1      & 64 x 64 x 16   & 63 x 63 x 32   \\ \cline{2-6}
                   & Max pooling               & -                      & 2      & 63 x 63 x 32   & 31 x 31 x 32   \\\hline
\multirow{2}{*}{3} & Convolution + RELU        & 64                     & 1      & 31 x 31 x 32   & 30 x 30 x 64   \\ \cline{2-6}
                   & Max pooling               & -                      & 2      & 30 x 30 x 64   & 15 x 15 x 64   \\\hline
\multirow{2}{*}{4} & Convolution + RELU        & 128                    & 1      & 15 x 15 x 64   & 14 x 14 x 128  \\ \cline{2-6}
                   & Max pooling               & -                      & 2      & 14 x 14 x 128  & 7 x 7 x 128    \\\hline
5                  & Fully Connected + RELU    & 100                    & -      & 6272           & 100            \\ \hline
6                  & Fully Connected + Sigmoid & 2                      & -      & 100            & 2             \\\hline

\end{tabular}
\caption{Architectural details of the custom 2D-CNN model.}
\end{table*}

\section{Experimental Results}

\subsection{Performance Metrics}

The performance of the proposed 2D-CNN classifier is evaluated by computing the following three performance metrics: accuracy, sensitivity, and specificity, as follows:
\begin{align}
    Accuracy (\%)=\frac{(TP+TN)}{(TP+TN+FP+FN)}  \times 100   
\end{align}
\begin{align}
    Sensitivity (\%)=\frac{TP}{(TP+FN)}  \times 100 
\end{align}
\begin{align}
    Specificity (\%)=\frac{TN}{(TN+FP)}  \times 100  
\end{align}

In the above equations, TP stands for true positive (an ECG beat correctly classified as belonging to MI class), TN stands for true negative (an ECG beats correctly classified as belonging to the healthy class), FP stands for a false positive, (miss-classification of a healthy ECG beat as MI), and FN stand for false-negative (miss-classification of an MI ECG beat as healthy).

\subsection{Experiment Details \& Hyper-parameters}

Next, the implementation details, hyper-parameters, and experimental results achieved by the proposed approach are discussed. As described earlier, ECG beats in the PTB dataset were transformed into four types of images/datasets: namely, DS1 (GASF images with noise and baseline wander), DS2 (GADF images with noise and baseline wander), DS3 (GASF images with noise and baseline wander removed), and DS4 (GADF images with noise and baseline wander removed). For pre-processing and implementation of the custom 2D-CNN, we used Python environment with Keras and Tensor-flow deep learning libraries. For training and testing of the custom 2D-CNN, an 80-20 training/test data split was used. Additionally, we used the K-fold cross-validation approach (with K=10) to prevent over-fitting \cite{duda33}. Thus, the overall classification accuracy for each dataset was obtained as the average value of the accuracy recorded in all 10 cycles of K-fold cross-validation. The accuracy and loss curves became saturated/stable in less than 50 epochs during both training and validation phase for all datasets (DS1, DS2, DS3 and DS4), as can be seen in Fig. 13 and Fig. 14. We used "Adam" optimizer and "cross-entropy" as the loss function. We used gradient-descent method with a learning rate of 0.001, and fixed the batch size to 8. 

\begin{figure}[h]
    \centering
    \includegraphics[width=0.5\textwidth]{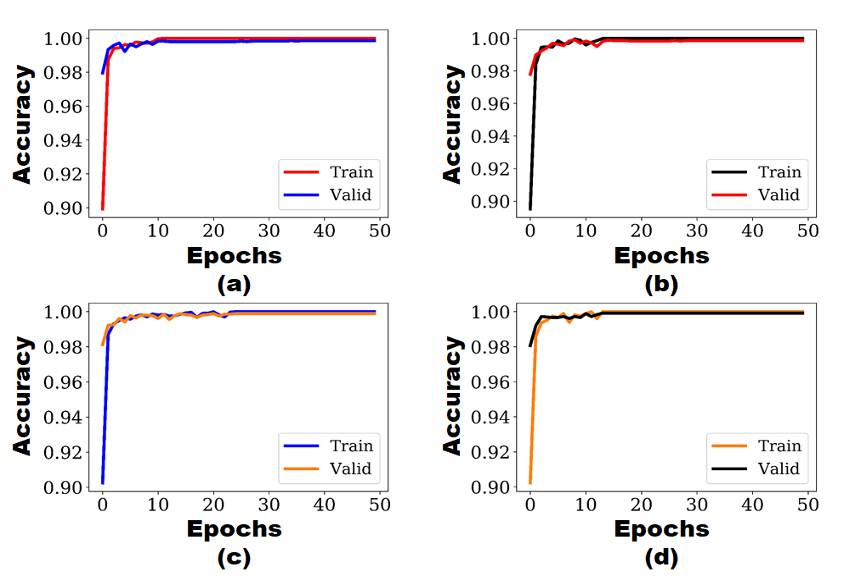}
    \caption{Accuracy curves for four datasets (a) GASF method with noise and baseline wander, (b) GADF method with noise baseline wander, (c) GASF method without noise and baseline wander, and (d) GADF method without noise and baseline wander. }
    \label{fig:mesh1}
\end{figure}

\begin{figure}[h]
    \centering
    \includegraphics[width=0.5\textwidth]{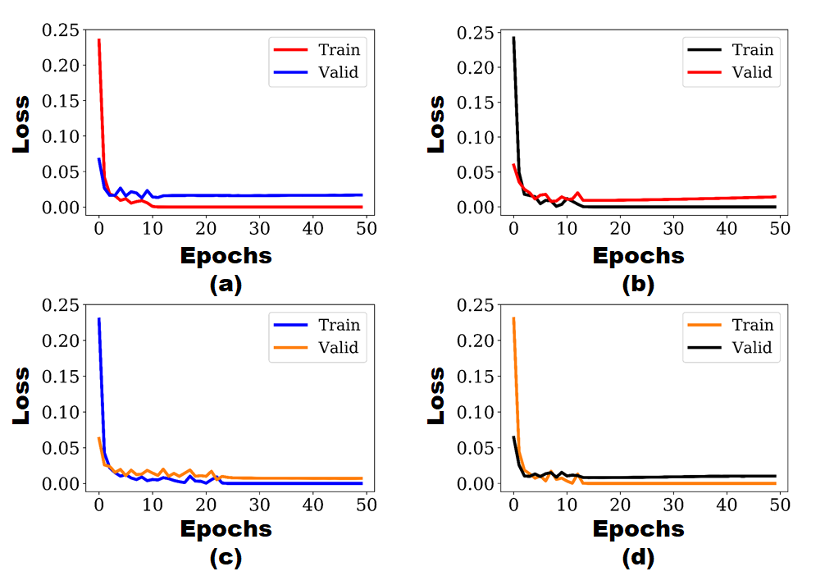}
    \caption{Loss curves for four datasets (a) GASF method with noise and baseline wander, (b) GADF method with noise and baseline wander, (c) GASF method without noise and baseline wander, and (d) GADF method without noise and baseline wander. }
    \label{fig:mesh1}
\end{figure}

\subsection{Results} 
Our proposed method achieved an average classification accuracy of 99.68\%, 99.80\%, 99.82\%, and 99.84\% when utilizing DS1 (GASF images with noise and baseline wander), DS2 (GADF images with noise and baseline wander), DS3 (GASF images with noise and baseline wander removed), and DS4 (GADF images with noise and baseline wander removed). More detailed results (i.e., performance metrics of accuracy, sensitivity, and specificity) for the best classification result achieved for each dataset are summarized in Table III.

\begin{table*}[]
\begin{tabular}{|c|c|c|c|c|c|c|c|c|c|c|c|} \hline
\textbf{Model Type} & \textbf{With noise and BW} & \textbf{Without noise and BW} & \textbf{GASF} & \textbf{GADF} & \textbf{\begin{tabular}[c]{@{}c@{}}TN\\ Beats\end{tabular}} & \textbf{\begin{tabular}[c]{@{}c@{}}TP\\ Beats\end{tabular}} & \textbf{\begin{tabular}[c]{@{}c@{}}FP\\ Beats\end{tabular}} & \textbf{\begin{tabular}[c]{@{}c@{}}FN\\ Beats\end{tabular}} & \textbf{\begin{tabular}[c]{@{}c@{}}ACC\\ (\%)\end{tabular}} & \textbf{\begin{tabular}[c]{@{}c@{}}SEN\\ (\%)\end{tabular}} & \textbf{\begin{tabular}[c]{@{}c@{}}SPE\\ (\%)\end{tabular}} \\ \hline
DS 1                & Y                             & N                                & Y             & N             & 2505                                                        & 7530                                                        & 19                                                          & 13                                                          & 99.68                                                       & 99.8                                                        & 99.2                                                        \\ \hline
DS 2                & Y                             & N                                & N             & Y             & 2508                                                        & 7539                                                        & 16                                                          & 4                                                           & 99.80                                                       & 99.9                                                        & 99.3                                                        \\ \hline
DS 3                & N                             & Y                                & Y             & N             & 2509                                                        & 7540                                                        & 15                                                          & 3                                                           & 99.82                                                       & 99.9                                                        & 99.4                                                        \\ \hline
DS 4                & N                             & Y                                & N             & Y             & 2510                                                        & 7541                                                        & 14                                                          & 2                                                           & 99.84                                                       & 99.7                                                        & 99.4                                                       \\ \hline
\end{tabular}
\caption{Classification results for all four datasets.
(DS=Data set, TN=True Negative, TP=True Positive, FP=False Positive, FN=False Negative, ACC=Accuracy, SEN=Sensitivity, SPE=Specificity, BW=baseline wander).
}
\end{table*}

\subsection{Discussion}

Fig. 15 compares the performance achieved by the proposed method with the state-of-the-art MI detection works. Many of the existing works use 12-lead ECG signals to classify between the healthy and MI ECG beats (12-lead ECG provides more data to the classifier, and thus, helps it achieve a superior performance; but at the same time, it also increases the complexity of the classifier by a big amount). In contrast to many of the existing methods, our proposed method utilizes only the lead-II of the ECG for MI detection. Fig. 15 demonstrates that, when compared with previous works, the proposed method provides excellent classification accuracy, even when trained on noisy ECG signals with baseline wander intact. 

\begin{figure*}
  \includegraphics[width=0.9\textwidth,height=11cm]{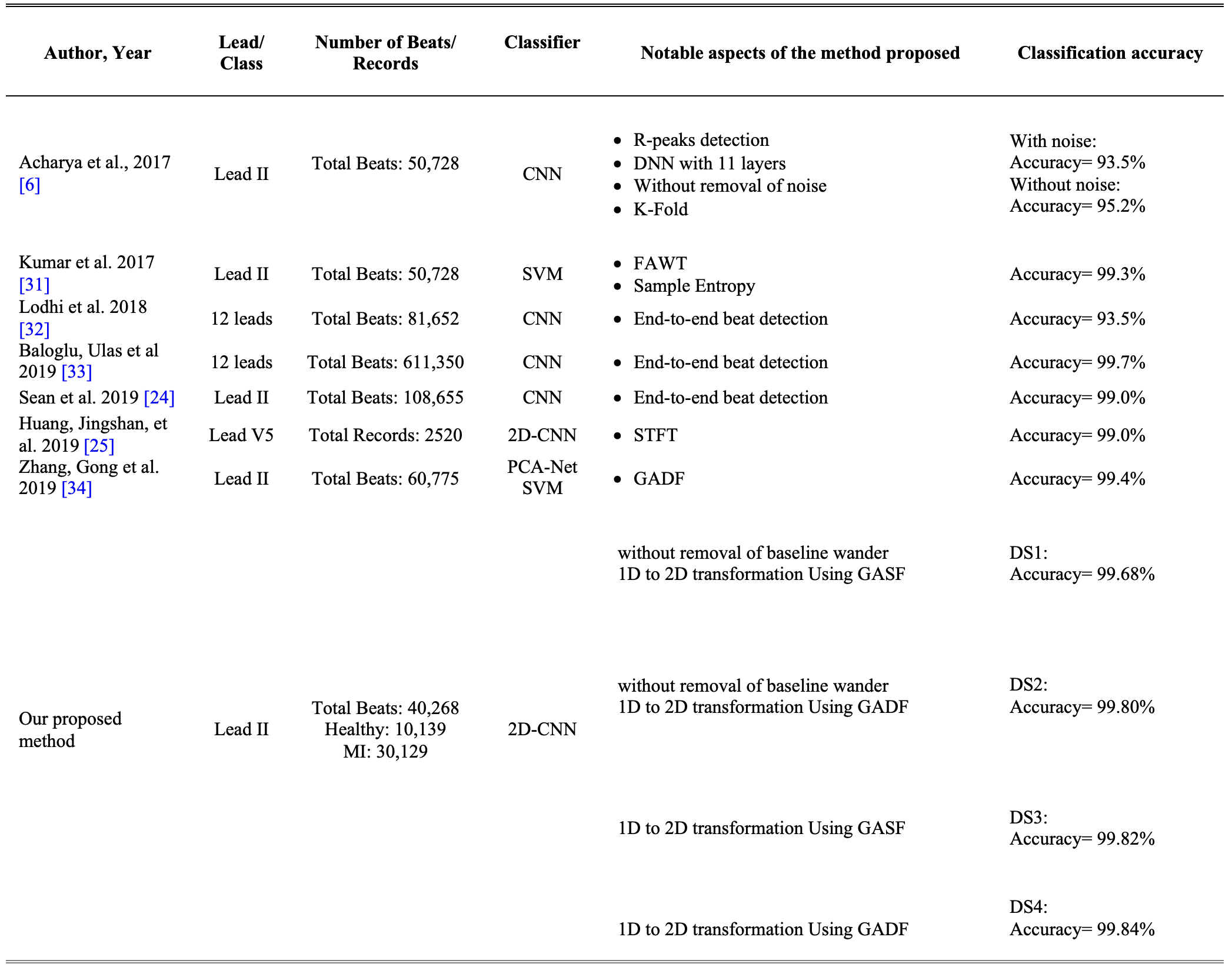}
  \caption{A comprehensive comparison between our proposed method and the state-of-the-art: [6],[24],[25],\cite{kumar39},\cite{lodhi40},\cite{baloglu41},\cite{zhang42}.}
\end{figure*}

\section{Conclusion}

This paper proposed a novel framework for automated inferior MI detection using lead-II of ECG. We utilized the PTB dataset from Physionet to train and test our custom 2D-CNN classifier. Specifically, we utilized the Gramian angular field (under summation and difference) method to transform the ECG data into gray-scale images in order to feed them into the 2D-CNN (with six hidden layers) which classified the ECG beats as belonging to either healthy subjects or subjects with MI. The proposed 2D-CNN model achieved a high classification accuracy that lies in the range 99.68
Future work will look into the following: 1) incorporation of other ECG leads (leads I, V5), in addition to lead II, for improved MI detection and localization, 2) design and evaluation of domain-adaptation methods using other ECG arrhythmia datasets (with different population groups) in order to develop generalizable and robust AI methods for MI detection, 3) alternate time-frequency analysis methods for ECG signal analysis for MI detection.

\ifCLASSOPTIONcaptionsoff
  \newpage
\fi



%

\vspace{-0.2cm}
\footnotesize{
\bibliographystyle{IEEEtran}
\bibliography{anystyle}
}

\vfill\break

%








\end{document}